\documentclass[twoside]{article}
\usepackage{fleqn,espcrc2}
\usepackage{graphicx}

\title{Laser Cooling of TeV Muons\thanks{Contributed to the NuFAct'00 
International Workshop.}
\hfill Fermilab-Conf-00/162}

\author{Fritz DeJongh\address{Fermi National Accelerator 
Laboratory, fritzd@fnal.gov}}

\begin{document}

\begin{abstract}
We show that Compton scattering can be used to cool
TeV-scale muon beams,
and we derive analytical expressions for the equilibrium transverse angular
spread, longitudinal energy spread, and power requirements.  We find that
a factor of a few thousand reduction in emittance is possible for a
3 TeV muon collider.
\end{abstract}

\maketitle

\section{Introduction}

Muon colliders are a possible future tool for the exploration of
physics at the TeV scale and beyond.  The current status of the 
developent of the muon collider concept is described in Ref.~\cite{MCC},
which includes a description of a 3 TeV center-of-mass (COM) machine.

One of the major challenges to realizing a high-luminosity muon collider
is to cool the diffuse bunches of muons produced in pion decays.  Much
effort is being put into the problem of quickly
cooling these bunches at the
front-end of a muon-collider and maintaining the low emittance while
bunches are being accelerated and brought into collision.  

For the 3 TeV COM machine, the problems of neutrino radiation~\cite{rad}
and power consumption are already becoming prohibitive.
Additional 
cooling of the bunches after acceleration would mitigate these problems
by allowing a given luminosity to be attained with
fewer stored muons and a lower repitition rate, and allow consideration
of even higher energy muon colliders.  For example, the tunnels and
high-field magnets being discussed for future hadron colliders~\cite{VLHC} 
could ultimately be used for
a 100 TeV-scale muon collider.
Post-acceleration cooling would also reduce detector backgrounds
from muon decay.

The possibility of using Compton scattering for 
cooling of electron bunches for
$\gamma \gamma$ colliders has been previously considered~\cite{gg}.
The luminosity for
$e^+ e^-$ collisions is already limited by beamstrahlung effects,
but additional cooling may greatly improve the $\gamma \gamma$
luminosity.

We propose herein the possibility of post-acceleration cooling 
of muons beams using Compton scattering.

\section{Compton Scattering}

The Compton Scattering cross-section, in the rest frame of the muon, is:
\begin{equation}
\frac{d\sigma}{d\Omega} = \frac{\alpha^2}{2m^2}
\left( \frac{k'}{k} \right)^2
\left( \frac{k'}{k} + \frac{k}{k'} - \sin^2 \theta \right),
\end{equation}
where
\begin{equation}
k' = \frac{k}
          {1 + (k/m)(1-\cos\theta)}
\end{equation}
and $k$ is the incoming photon energy, $k'$ is the outgoing
photon energy, and $\theta$ is the photon scattering angle.
For $k \ll m$, $k' \approx k$ and the scattering is
roughly isotropic:
\begin{equation}
 \frac{d\sigma}{d\Omega} \propto (1+\cos^2 \theta), 
\end{equation}
and the total cross-section
is given by:
\begin{equation}
\sigma_{\rm C}  =  \frac{8\pi}{3} \frac{\alpha^2}{m^2}
                =  \frac{8\pi}{3} r^2_\mu
\end{equation}
where $r^2_\mu$ is the classical muon radius.
Thus, compared to electrons, the cross-section is reduced by 
$\approx 4\times 10^4$.  

As an example, to expect 1 collision with a 0.1 eV photon, the light
energy density would need to be 10 $\rm J/(\mu m)^2$.

Consider a beam of muons, with energy and momentum defined by
$\beta$, $\gamma$, $E_\mu$, and $p_\mu$, colliding head-on with
a mono-energetic beam of photons with energy $E_\gamma$.
We will approximate $\beta = 1$.
The energy of the photons in the muon rest frame is:
\begin{equation}
E^*_\gamma = \gamma E_\gamma (1 + \beta) \approx 2\gamma E_\gamma.
\end{equation}
On average, the photon will transfer 
longitudinal momentum $E^*_\gamma$ to the muon.  In the lab frame,
\begin{equation}
E_\mu \to E_\mu - 2 \gamma^2 E_\gamma.
\end{equation}
Typically, the muon will receive a smaller transverse kick:
\begin{equation}
p_{T\mu} \approx \gamma E_\gamma.
\end{equation}
Therefore, for large $\gamma$, the muon is essentially slowed down
without changing direction.

If lower photon energies are needed, this can effectively be
achieved by aiming the photon beam at an angle $\theta$ relative to
the head-on direction.  The transverse momentum of the photon is
inconsequential.  The effective photon energy becomes:
\begin{equation}
E_\gamma \to E_\gamma (1+\cos\theta) / 2 .
\end{equation}

\section{Cooling Effect}

For simplicity,
we consider the case that the the muons undergo on average one
Compton scattering, and are afterwards reaccelerated to compensate
for the average energy loss.  The following conclusions are also valid for
any average number of scatterings, as long as the relative energy loss
is small.

Transversely, a muon has a small angle
$\alpha$ relative to the beam direction in a plane, for example the
$x - z$ plane.
After the scattering,
$\alpha$ remains unchanged, within an amount $E_\gamma / m$.
After reacceleration,
\begin{equation}
\alpha \to \alpha ( 1 - 2 \gamma E_\gamma / m ).
\end{equation}
Thus, the angular spread can be reduced, 
in the same way as for ionization cooling at low energy.

There is also a heating effect from the spread in the transverse kick.
The average increase in the variance of $\alpha$ is equal to that for
one Compton scattering:
\begin{equation}
\sigma^2_\alpha \to \sigma^2_\alpha + \frac{6}{5}
    \left(\frac{E_\gamma} { m} \right)^2 .
\end{equation}
To find the equilibrium angular spread we equate the cooling and
heating effects to find:
\begin{equation}
\sigma_\alpha = \sqrt{ \frac{3}{10} \frac{E_\gamma}{E_\mu} }.
\label{trans}
\end{equation}

Luminosity is inversely proportional to the sums of the emittances
of the two colliding
beams.  It is also proportional to the product of the numbers
of muons in the two beams.  This product decays with a time constant
of one-half the muon lifetime.
For a fractionally small energy loss per collision, the number
of Compton scatterings needed to reduce $\sigma_\alpha$ by
a factor of $1/e$ is given by:
\begin{equation}
n = \frac{m}{2\gamma E_\gamma}
\end{equation}
The total energy used to reaccelerate the muon after these $n$ collisions
is equal the original muon energy:
\begin{equation}
E_{reacc} = E_\mu \label{reacc} . 
\end{equation}
The power density needed to attain this factor in one-half muon lifetime
is given by:
\begin{equation}
p  =  \frac{n E_\gamma}{\sigma_{\rm C} \gamma \tau_\mu / 2 } 
   =  \frac{3}{8\pi} \frac{m}{\gamma^2 \tau_\mu r_\mu^2} \label{power} .
\end{equation}
Thus, the power required decreases as the square of the muon energy
increases.

Longitudinally, since the average energy loss in a Compton scattering is 
greater the higher the muon energy,
there is an energy bunching effect.
For $\sigma_{E_\mu} \ll E_\mu$, and the average case of one Compton scattering,
the bunching effect is:
\begin{equation}
\sigma_{E_\mu} \to \sigma_{E_\mu} - 
\frac{\sigma_{E_\mu}}{E_\mu} 4\gamma^2 E_\gamma .
\end{equation}
There are also two sources of energy heating.  The first is from the 
variance in the number of Compton scatterings of the muon, given
by Poisson statistics.  This leads to a variance in the energy
spread of $(2\gamma^2 E_\gamma)^2$.  The second
is from the variance in the energy spread within one Compton scattering,
given by $2/5 \ (2\gamma^2 E_\gamma)^2$.
The total heating effect is then:
\begin{equation}
\sigma_{E_\mu}^2 \to \sigma_{E_\mu}^2 +
     (2 \gamma^2 E_\gamma)^2 + \frac{2}{5} (2 \gamma^2 E_\gamma)^2 .
\end{equation}
Equating the heating and cooling effects, we find for the
equilibrium energy spread:
\begin{equation}
\frac{\sigma_{E_\mu}}{E_\mu} = \sqrt{
\frac{7}{10} \frac{E_\mu E_\gamma}{m^2} } .
\label{long}
\end{equation}

We have checked these derivations with a simple Monte Carlo simulation
of a set of muons undergoing repeated Compton scatterings and boosts.
The predictions of these equations are in excellent quantitative agreement
with the simulation.

\section{Power Considerations}

In principle, Compton scattering could be used to cool low-energy
muons.  Unfortunately, our estimates show that the power requirements
would be prohibitive by many orders of magnitude.  However, as shown
in Eq.~\ref{power}, the power density needed decreases as the square of the 
muon energy increases, and may be reasonable at TeV energies.  

Two other general considerations affect how the power needed scales
with the muon energy.
First, the muons will be in a storage ring.  The photon pulses can
be placed in a cavity and reused once per turn of the muons around the ring,
as illustrated in Fig.~\ref{3tev}.  The size of the ring, and therefore
the time per turn, is proportional to the muon energy.  For a lower muon
energy, the photon pulses can be reused at a faster rate.  Therefore,
for this scheme, the power needed to produce the photons scales as
only $1/E$.

The second consideration is that the 
geometric emittance, and therefore spot size,
decreases linearly as the muon bunch is accelerated.  Therefore, the
area that needs to be illuminated, and the total power,
decreases as the square of the muon energy.
Putting these two considerations together, the power needed for cooling
scales as $1/E^3$.

The length of the muon bunch can also affect the amount of power needed.
A laser beam can be focussed to collide with the muon bunch.  The 
diffraction-limited spot
size at the focus is proportional to the F-stop ($F$).  However, the
depth of focus is proportional to $F^2$.  If the length of the muon
bunch is too long, we will need to increase the depth of focus, which
implies increasing the spot size, which will require proportionally more
total power.

Finally, Eq.~\ref{power} gives only the cooling rate needed to compensate
the luminosity
for the muon decays.  A cooling rate several times higher than this
will be necessary to realize large increases in the luminosity.
We also note that as cooling power is added, we also need to add RF
power to reaccelerate the muons, as described in Eq.~\ref{reacc}.

Although a lot of additional power will be needed for the laser and RF
systems, a much lower repitition rate will be needed for a given
luminosity, and thus the power consumption for the facility as a whole
may be much lower.

\section{Application to the 3 TeV Muon Collider}

\begin{figure}[t]
\includegraphics[width=3in]{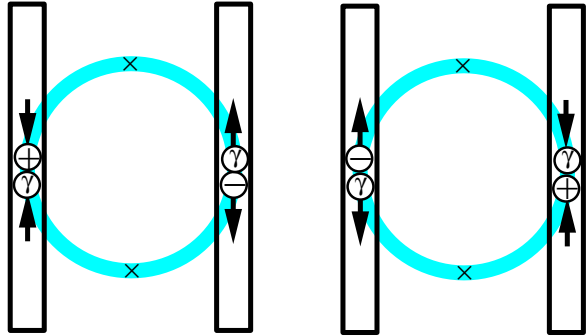}
\caption{Configuration of resonant cavities and muon storage ring.
On the left, each muon bunch collides head-on with a photon pulse.
On the right, one-half cycle later, each photon pulse has reflected
and collides head-on with the other muon bunch.}
\label{3tev}
\end{figure}

\begin{table*}[htb]
\caption{Parameters of the 3 TeV COM muon collider and possible laser
cooling systems.  These parameters assume that the full emittance
improvement will be attained in one luminosity lifetime.}
\label{table:1}
\newcommand{\m}{\hphantom{$-$}}
\newcommand{\cc}[1]{\multicolumn{1}{c}{#1}}
\renewcommand{\tabcolsep}{2pc} 
\renewcommand{\arraystretch}{1.2} 
\begin{tabular}{@{}llll}
\hline
	& Ref.~\cite{MCC} & $ \rm CO_{2}$ laser & Nd:Glass Laser \\
\hline
Bunches/fill	& 4 	& 1 	& 1 \\
Rep. rate (Hz)	& 15	& 1	& 1 \\
Initial beam width ($\mu$m)	& 3.2	& 3.2	& 3.2 \\
Initial bunch length (mm)  	& 3.0	& 1.0	& 1.0 \\
$\lambda$ ($\mu$m)	&	& 10.6	& 1.05 \\
$E_\gamma$ (eV)	& 	& 0.1	& 1.0	\\
Emittance improvement	&	& 7000	& 2200 \\
Energy spread (\%)
	& 0.16	& 0.3	& 1.0 \\
F-stop	&	& 7	& 22	\\
Diffraction-limited width ($\mu m$) &	& 43	& 13 \\
Energy in photons (MJ) &	& 4	& 0.4	\\
Efficiency (\%) &
	& 25	& 1 \\
$\mu$/bunch	& $2 \times 10^{12}$	& $3 \times 10^{11}$	&
		$5 \times 10^{11}$ \\
Tune shift	& 0.044	& 31	& 15 \\
Luminosity ($\rm cm^{-2} s^{-1}$)	&
	$7 \times 10^{34}$ & $7 \times 10^{34}$ & $7 \times 10^{34}$ \\
\hline
\end{tabular}\\[2pt]
\label{tab:laser}
\end{table*}

We start with the parameters for the 3 TeV muon collider
in Ref.~\cite{MCC}.
The
ring has a circumference of 6 km, and contains 4 bunches at a time.
We assume that this will be reduced to 1 bunch.  We have assumed that
the bunch length can be shortened to 1 mm.
Midway between
each crossing point, we place a resonant 
optical cavity.  
This is illustrated in Fig.~\ref{3tev}.
Each cavity contains
a photon pulse reflecting back and forth.  The length of the cavity
is set to 3 km, so that the light pulse hits a muon bunch in 
alternating
directions once per reflection.  
The luminosity lifetime is 15 msec, or 750 turns.  Thus, the cavity
should have a Q of $\approx 10^3$.  The length of the photon pulse should
be comparable to the $\beta^*$ of the machine, or about 30 ps.

Progress in improving laser intensity and decreasing photon pulse widths
has been very rapid~\cite{Mourou}.  For example, the Mercury 
project~\cite{Mercury} is developing a
1.05 $\mu$m laser system that will generate 100 J at 10 Hz in 5 ns pulses
at 10\% efficiency.
We have considered two types of laser systems
to generate the necessary photon pulses:  $\rm CO_{2}$ lasers,
which typically have a 10.6 $\mu$m wavelength and 25\%
efficiency, and Nd:Glass lasers, which typically have a 1.05 $\mu$m
wavelength and 1\% efficiency.
The parameters of these laser systems with the 3 TeV COM machine
are shown in Table~\ref{tab:laser}.

As shown in Table~\ref{tab:laser}, we can expect emittance
improvements by a factor of a few thousand.  However, the energy
needed in the laser pulses is very high:  Of order one MJ.  The two
laser cooling stations in Fig.~\ref{3tev} would probably need to be
divided into several cooling stations each with a fraction of the
laser energy.  Shortening the muon bunch length could also allow
a considerable reduction in laser energy.  
It may be possible to exploit the reduction in beam size that occurs
during the cooling process.
Finally, a higher energy
muon collider would also reduce the laser energy requirement.

Some of the other parameters of the collider with laser cooling
may present challenges for the machine design.
The small emittance leads to a very high value of the tune shift.
Also, the energy spreads are somewhat high, especially for the
lower-wavelength laser.

\section{Conclusions}

We have shown that Compton scattering can be used to cool muon beams.
Eq.~\ref{trans} and Eq.~\ref{long} describe the achievable transverse angular
spread and longitudinal energy spread.

While the power needed for laser cooling of muon beams is very high,
it decreases as the cube of the muon energy increases, and may become
practical by TeV energies.  If so, emittance reductions by factors
of a few thousand are possible.  This would allow a given luminosity
to be attained with a much lower repitition rate, much less detector
background from muon decays, and a much reduced neutrino radiation hazard.
Several challenges remain to develop a plausible optical, laser, and machine
system.

\section{Acknowledgements}

Thanks to the organizers of the NuFact'00 International Workshop for the
opportunity to present these results.  Thanks to David Neuffer for several
useful conversations.

This work was performed at the Fermi National Accelerator Laboratory,
which is operated by Universities Research Association, under contract
DE-AC02-76CII03000 with the U.S. Department of Energy.

\end{document}